# VFLAIR-LLM: A Comprehensive Framework and Benchmark for Split Learning of LLMs

Zixuan Gu
School of Software, Tsinghua University
Beijing, China
gu-zx24@mails.tsinghua.edu.cn

Qiufeng Fan
Wuxi Innovation Center of Tsinghua AIR
Wuxi, China
fan.qiufeng@u.nus.edu

Long Sun
Wuxi Innovation Center of Tsinghua AIR
Wuxi, China
cnlonger@gmail.com

Yang Liu*
the Hong Kong Polytechnic University
Hong Kong, China
yang-veronica.liu@polyu.edu.hk

Xiaojun Ye
School of Software, Tsinghua University
Beijing, China
yexj@tsinghua.edu.cn

## Abstract

With the advancement of Large Language Models (LLMs), LLM applications have expanded into a growing number of fields. However, users with data privacy concerns face limitations in directly utilizing LLM APIs, while private deployments incur significant computational demands. This creates a substantial challenge in achieving secure LLM adaptation under constrained local resources. To address this issue, collaborative learning methods, such as Split Learning (SL), offer a resource-efficient and privacy-preserving solution for adapting LLMs to private domains. In this study, we introduce **VFLAIR-LLM** (available at https://github.com/FLAIR-THU/VFLAIR-LLM), an extensible and lightweight split learning framework for LLMs, enabling privacy-preserving LLM inference and fine-tuning in resource-constrained environments. Our library provides two LLM partition settings, supporting three task types and 18 datasets. In addition, we provide standard modules for implementing and evaluating attacks and defenses. We benchmark 5 attacks and 9 defenses under various Split Learning for LLM(SL-LLM) settings, offering concrete insights and recommendations on the choice of model partition configurations, defense strategies, and relevant hyperparameters for real-world applications.

*Corresponding author, also affiliated with the Shanghai Artificial Intelligence Laboratory.

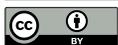

## 1 Introduction

The recent development and success of large language models (LLMs) have significantly reshaped the landscape of artificial intelligence, showcasing exceptional capabilities across a wide range of tasks. LLM training relies heavily on massive, high-quality data, fueling growing demand for such resources. Public data sources, such as books, web crawls, and open-access articles, have historically served as the backbone of LLM training data. However, research [38] indicates that the availability of public human text data is nearing exhaustion. This growing data scarcity has emerged as a critical bottleneck for LLM development, compelling a shift toward leveraging private domain data, which subsequently raises significant privacy concerns.

Sensitive data within private domains cannot be freely shared or processed by external LLM systems due to risks of data breaches, regulatory violations, and potential misuse. These challenges make the direct integration of private data into LLM training impractical. One potential solution is the local deployment of LLMs. However, this method requires substantial local computational resources, posing a significant barrier for smaller organizations or individuals managing sensitive private data. To address this challenge of private adaptation of LLMs under constrained local resources, various methods have been proposed. Off-site tuning[42] and knowledge distillation[13] leverage compact language models to approximate the behavior of target LLMs. However, these methods often suffer from notable performance degradation and require complex algorithmic implementations.

An alternative approach is Split Learning (SL)[10, 37], a collaborative training paradigm developed based on Federated Learning(FL) [22, 43]. It introduces a cross-silo scenario where a model is partitioned across participants, offering the benefit of minor performance degradation and a simple yet effective algorithmic implementation. However, this solution still faces considerable privacy concerns[3], as the server may attempt to infer clients' local data through various privacy attacks[45]. Various defense methods[7, 24] have also been proposed to address these risks.

In this work, we focus on leveraging SL for the private adaptation of LLMs. Aiming to support relevant research and applications, we design a lightweight and highly extensible Split Learning LLM framework, named VFLAIR-LLM.

VFLAIR-LLM incorporates basic modules for customizable SL-LLM inference and fine-tuning, including user-defined LLM partition, defense strategies, and other relevant functions. It supports 3 types of LLM architect and 3 corresponding task types, each with relevant datasets available for direct usage, and is open to users to add new datasets. To enable flexible privacy assessment and algorithm development, VFLAIR-LLM also offers multiple attack and defense methods in a modular style, ensuring easy usage and extension. In summary, our contributions are listed below:

- We develop a lightweight framework named VFLAIR-LLM for split learning of LLMs. This framework incorporates an easily adaptable model partition method for a wide variety of LLMs. Additionally, it addresses possible privacy concerns featuring 3 model inversion attacks(MIA), 2 label inference attacks(LIA), and 9 defenses.
- We conduct a comprehensive benchmark on attacks and defenses within the SL-LLM setting using VFLAIR-LLM. The benchmark provides various recommendations and insights on model partition configuration, defense strategies, and relevant hyperparameter selection to facilitate easy usage.

## 2 Related Works

### 2.1 Private adaptation of LLMs under limited local resources

Various methods have been proposed to address the challenges of private adaptation of LLMs under constrained local resources. Offsite-Tuning[42, 44] and knowledge distillationHsieh et al. [13] focus on training smaller, task-specific models locally to emulate the traditional LLM adaptation process. However, they often face trade-offs in model performance due to the inherent limitations of smaller models. Another possible solution to this concern is Split Learning(SL)[11, 34], an evolution of Federated Learning (FL) [22] where the model is partitioned across collaborators. In this approach, participants collaboratively train an LLM by exchanging model intermediate and gradients, allowing the data holder to train only a small portion of the full LLM. Various projects and frameworks [41, 50, 52] have been developed to facilitate research and deployment in this area. For example, VFLAIR [52] is an open-sourced library that supports SL training with a wide range of models, datasets, and protocols.

### 2.2 Split Learning of LLMs

As summarized in Table 1, several studies have explored SL for fine-tuning and inference of LLMs. SAP[30] is a privacy-preserving federated fine-tuning framework where a LLM is divided into 2 parts: "head" and "tail"(termed **HT** in the following discussion), aiming to defend model inversion attacks. Also leveraging a "head-tail" partitioning, SplitLoRA[19] introduced a fine-tuning framework for SL-LLM, demonstrating superior training performance. [3] proposed SplitLLM, which partitions the model into 3 parts: "head", "body" and "tail"(termed **HBT** in the following discussion). It introduces a novel data reconstruction attack(BiSR, tested in our following evaluations) to invert data input, highlighting the potential privacy risks in SL-LLM. While these efforts have laid the groundwork for SL-LLM research, they primarily focus on specific methodologies, with limited attention given to comprehensive privacy benchmarking and the development of user-friendly, extensible tools for broader adoption. This gap inspired us to develop a framework that not only provides comprehensive privacy algorithms but also prioritizes ease of use and extensibility.

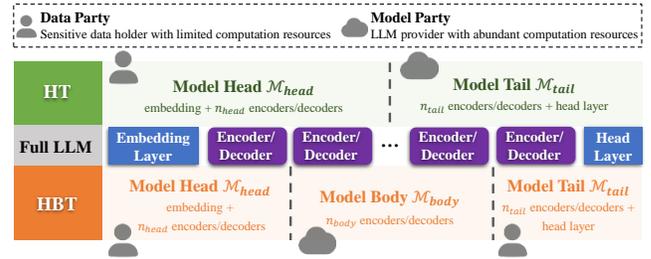

Figure 1: LLM Partition

## 3 Framework Overview

### 3.1 Split Learning for LLM(SL-LLM)

VFLAIR-LLM proposes a general Split LLM framework with a Data Party and a Model Party. ***Data party*** simulates participants equipped with data and labels but constrained computational resources for comprehensive LLM utilization, possessing only a few layers of a complete LLM. Meanwhile, the ***Model party*** simulates the LLM provider, retaining the majority segment of the LLM.

*3.1.1 LLM Partition in VFLAIR-LLM.* As described in Figure 1, we propose 2 SL-LLM settings: **"Head-Tail"(HT) SL-LLM** and **"Head-Body-Tail"(HBT) SL-LLM**, depending on how the LLM model is partitioned among parties, For encoder-only and decoder-only LLMs, the model is split at customizable points within the encoder or decoder sequence for both HT and HBT settings. For encoder-decoder LLMs, HBT splits the model at the encoder sequence and the decoder sequence. While HT partitions only the encoder sequence.

*3.1.2 Head-Tail(HT) SL-LLM.* In HT SL-LLM, a full LLM with $n = n_{head} + n_{tail}$ encoders/decoders is separated into a Model Head and a Model Body[19, 30]. Model Head $\mathcal{M}_{head}$ is allocated to the Data Party, containing the embedding layer and $n_{head}$ encoders/decoders. Model Tail $\mathcal{M}_{tail}$ is the rest of LLM held by the Model Party, containing $n_{tail}$ encoders/decoders and a head layer. Typically, $n_{tail}$ is set significantly larger than $n_{tail}$.

During forward propagation, the data party performs forward propagation first. Intermediate $H_1 = M_{head}(X)$ is then transmitted to the model party for further generation of the final output $\hat{Y} = M_{tail}(H_1)$. Backward propagation is performed in a reversed order with the model party transmitting the intermediate's gradient $G_1$ back to the data party. The detailed training algorithm is described in Algorithm 1 and Figure 2a. Note that HT SL-LLM assumes the model party can access the inference results or labels. For scenarios where label and inference results need to be further protected from the Model party, we further introduce the HBT SL-LLM[3].

Table 1: Summary of SL-LLM frameworks.

|  | LLM Types | SL-LLM Partition | | Attack | | Defense | Fine-tuning Strategy | Work Mode | | Evaluation Metrics | | |
|---|---|---|---|---|---|---|---|---|---|---|---|---|
|  |  | Head-Tail | Head-Body-Tail | Model Inversion | Label Inference |  |  | Standalone | Distributed | Performance | Privacy | Efficiency |
| SAP[30] | 1 | ✓ |  | ✓ |  | 0 | ✓ | ✓ |  | ✓ | ✓ |  |
| SplitLoRA[19] | 1 | ✓ |  | ✓ | ✓ | 0 | ✓ | ✓ |  | ✓ |  | ✓ |
| SplitLLM[3] | 11 |  | ✓ | ✓ | ✓ | 3 | ✓ | ✓ |  | ✓ |  |  |
| VFLAIR-LLM | 16 | ✓ | ✓ | ✓ | ✓ | 9 | ✓ | ✓ | ✓ | ✓ | ✓ | ✓ |

*3.1.3 Head-Body-Tail(HBT) SL-LLM.* The HBT SL-LLM splits a full LLM with $n = n_{head} + n_{body} + n_{tail}$ encoders/decoders into 3 parts. Model Head $M_{head}$ contains the embedding layer and the first $n_{head}$ encoder/decoder layers, which is allocated to the Data Party. Model Body $M_{body}$ contains $n_{body}$ encoders/decoders, the main body part of LLM, and is allocated to the Model Party. Model Tail $M_{tail}$ contains $n_{tail}$ encoders/decoders and a head layer, which is allocated to the Data Party. By allocating both the model head and model tail to data party, this setting can hinder direct label inference and model output infringement by the model party. Typically, $n_{body}$ is set significantly larger than $n_{head}$ and $n_{tail}$.

During the forward process, the data party first feeds input data into its model head. Its intermediate $H_1 = M_{head}(X)$ is then transmitted to the model party for model body forward calculation: $H_2 = M_{body}(H_1)$. Finally, the model body output is transmitted back to data party to generate final predictions $\hat{Y}$ using the model tail. During the backward propagation, data party and model party consecutively calculate gradients $G_1$ and $G_2$ for its received intermediates and perform local backward calculations. Detailed training algorithm is described in Algorithm 2 and Figure 2b.

## 3.2 Fine-tuning Methods for SL-LLM

To enable efficient LLM fine-tuning, various fine-tuning strategies[14, 19] have been proposed. VFLAIR-LLM enables users to customize their own fine-tuning strategies, including **Full-Tuning**, where all model parameters are trainable, and **Local-Tuning**, where only the data party's sub-model is trainable. After specifying the trainable model segments, we also incorporate the PEFT Library[21] into VFLAIR-LLM, enabling support for a wide range of parameter-efficient fine-tuning (PEFT) methods, including LoRA[14], LoKr[15], AdaLoRA[48], and LoHa[15] etc. Parties can choose to apply either a **Vanilla** fine-tuning strategy or a **LoRA** strategy to their own model segments.

## 3.3 VFLAIR-LLM Framework Design

Based on the codebase of VFLAIR [52], a general framework for vertical federated learning, we develop VFLAIR-LLM, a framework specific for implementing and benchmarking SL-LLM scenarios, as illustrated in Figure 3. VFLAIR-LLM shares VFLAIR's configuration design, party loading module, and basic communication functions, but focuses on LLM-centered datasets and tasks, fine-tuning strategies, and attack and defense evaluations.

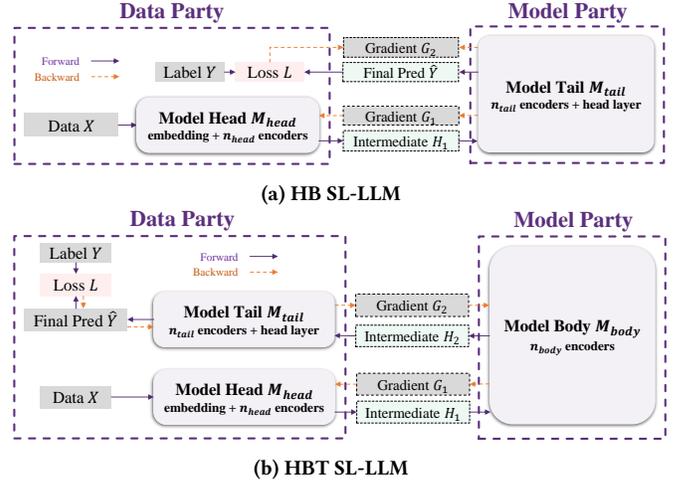

(a) HB SL-LLM

(b) HBT SL-LLM

Figure 2: Training Process of SL-LLM

*2 SL-LLM Partition Settings.* VFLAIR-LLM offers two LLM partition settings: Head-Body (HT) SL-LLM and Head-Body-Tail (HBT) SL-LLM as described in Section 3.1.

*2 Usage Pipelines.* VFLAIR-LLM supports both LLM fine-tuning and LLM inference. In *Inference* pipeline, users can load a pre-trained LLM to conduct direct inference on a given dataset. In *Fine-tune* pipeline, users can fine-tune an LLM on a downstream task.

*16 LLM Types.* Currently, we support 16 LLMs as shown in Table 2. To enable easy extension and compatibility, all model splits are implemented based on the Transformers [36] library with detailed guidance in our code base.

Table 2: Supported LLM Types

| Structure | LLM Types |
|---|---|
| Encoder-only | Bert Roberta Albert |
| Decoder-only | GPT2 Llama Baichuan2 ChatGLM2 Falcon Gemma Mamba Mistral Qwen2 Deepseek MiniCPM Qwen2-VL |
| Encoder-Decoder Qwen | T5 |

*3 Basic LLM Architects.* VFLAIR-LLM support 3 commonly used LLM architects as presented in Table 3, each featuring a different head layer added to the main LLM body to suit downstream tasks.

*2 Work Modes.* VFLAIR-LLM support 2 work modes: standalone simulation and distributed deployment, supporting both simulation research and real-world applications. We provide an efficiency comparison between distributed and standalone mode in Appendix C.

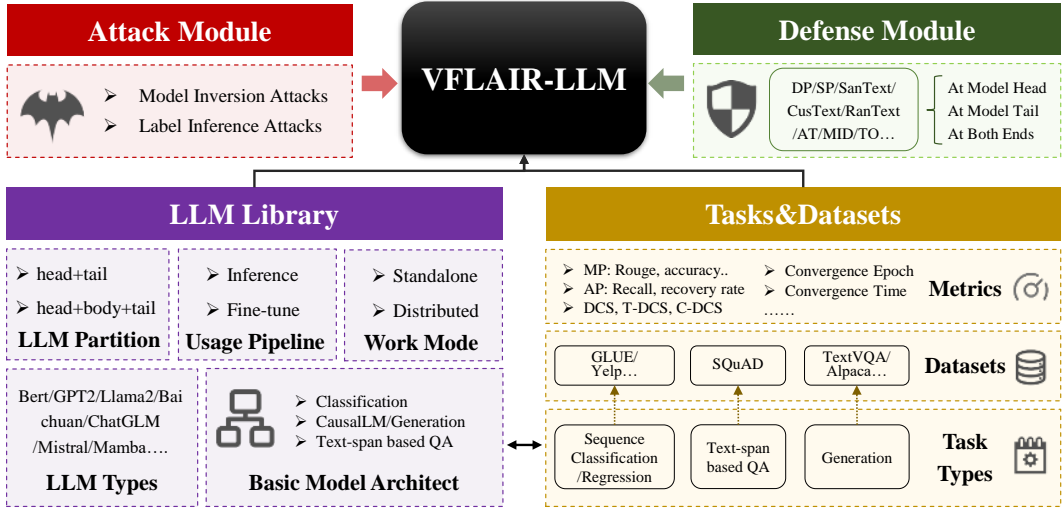

Figure 3: VFLAIR-LLM Framework Overview

Table 3: Supported Datasets and Tasks in SL-LLM

| Architect | Task Type | Dataset | Task Description |
|---|---|---|---|
| Classification (CLS) | Sequence Classification (Regression) | SST-2[39] | sentiment analysis |
| | | CoLA[39] | acceptability |
| | | MRPC[39] | paraphrase identification |
| | | QQP[39] | semantic equivalence |
| | | MNLI[39] | natural language inference |
| | | QNLI[39] | QA entailment |
| | | RTE[39] | textral entailment |
| | | WNLI[39] | pronoun resolution |
| | | Yelp[49] | review rating |
| | | STS-B[39] | semantic similarity |
| Text-span based QA (TQA) | Text-span based QA | SQuAD[27] | span-based question answering |
| CausalLM/ Generation (CLM) | Generation | Lambada [25] | next token prediction |
| | | Alpaca [33] | text generation |
| | | Dolly [6] | text generation |
| | | CodeAlpaca [2] | code generation |
| | | MATH [12] | math |
| | | GSM8K[5] | math |
| | | TextVQA [31] | visual question answering |

*Attacks.* VFLAIR-LLM supports 3 model inversion attacks and 2 label inference attacks as summarized in Table 4. Detailed attack deployment is presented in Figure 4.

(1) Threat Model: In this work, we assume the model party is an honest-but-curious attacker for both MIA and LIA. It follows the given SL-LLM protocol and does not collude with external entities. Furthermore, we operate under a white-box attack scenario, meaning the adversary, as the model provider, possesses complete knowledge of all model slice parameters, representing a significantly strong attack scenario. Unless otherwise specified, the attacker does not possess any auxiliary data or information on data parties' data.

Table 4: Summary of attacks in VFLAIR-LLM

| Model Inversion Attack | VMI [8],RMI [32],BiSR [3] |
|---|---|
| Label Inference Attack | BLI [53],NS [18] |

(2) Attack Methodology: In **Model Inversion Attacks(MIA)**[3], also known as Embedding Inversion Attacks(EIA)[16, 17, 23], the model party will try to infer data party's original text $X$ from the transferred intermediate $H_1$. *Vanilla Model Inversion(VMI)* [8] is a learning based model inversion attack featuring a 2-step data reconstruction process. First, the attacker infers the original input embedding $E(X')$ through optimization by minimizing loss between the calculated intermediate $H' = \mathcal{M}_1(E(X'))$ and the real intermediate $H$. Secondly, it recovers tokens from the inferred embedding by choosing the max cosine similarity between the embedding matrix $E$ and the inferred embedding $E(X')$, generating inferred text $X'$. *Relaxation-based Model Inversion(RMI)* [32] follows a similar 2-phase data reconstruction, but conducts relaxation on each token vector of the input sequence with a continuous variable $z$ for optimization in the first phase. *Bi-directional Semi-white-box Reconstruction (BiSR)* [3] incorporates a noise-aware pretraining phase for embedding initialization before proceeding with the traditional procedure of VMI. It has demonstrated strong attack performance across various LLMs, including BERT, GPT2, Llama2, ChatGLM, and Flan-T5.

In **Label Inference Attacks(LIA)**[18, 53], the model party attempts to infer the data party's label data $Y$ from the gradient $\mathbf{G}_2$ received during training. In *Batch-level Label Inference(BLI)*[53], the adversary trains an inversion model to invert label information from batch-level gradients. *Norm-based Scoring(NS)* [18] is implemented by calculating sample-level gradient norm values to identify positive/negative labels for binary classification tasks, since the norm of gradients for positive samples are generally larger than negative ones when data is unbalanced distributed.

While MIA can be injected at both training-time and inference-time, LIA is a training-time attack as it requires gradient information.

*Defenses.* VFLAIR-LLM supports 6 perturbation-based defenses and 3 learning-based defenses as summarized in Table 5. Details about specific defense methods are listed in Appendix A. As depicted in Figure 4, applying defenses **at the model head** mitigates MIA threats, while deployment **at the model tail** hinders LIA.

Table 5: Summary of defenses and tested hyper-parameters.

| Defense | Appliable Position | | Hyper-parameter Values |
|---|---|---|---|
|  | model head | model tail |  |
| DP | ✓ | ✓ | $\epsilon$ = 500, 100, 70, 50 |
| SP | ✓ | ✓ | $r$ = 95.0%, 96.0%, 97.0%, 98.0% |
| SanText | ✓ |  | $\epsilon$ = 5, 1, 0.1, 0.01 |
| CusText | ✓ |  | $\epsilon$ = 5, 1, 0.1, 0.01 |
| RanText | ✓ |  | $\epsilon$ = 30, 25, 20, 15, 10 |
| SnD | ✓ |  | $\eta$ = 1e5, 1e4, 1e3, 100, 10 |
| AT | ✓ | ✓ | $\lambda$ = 5.0, 1.0, 0.1, 0.01, 0.001 |
| MID | ✓ | ✓ | $\lambda$ = $1e^{-5}, 1e^{-4}, 1e^{-3}$, 0.01, 0.1, 0.5 |
| TO | ✓ |  | $n_{cluster}$ = 250, 200, 150, 100, 50 |

*hyper parameter values are listed from weakest to strongest defense here.

**Perturbation-based defenses** such as Differential Privacy(DP) [24], Sparsification(SP) [1, 9, 54] and Split-N-Denoise(SnD) [20] add noise to model intermediate or gradients to prevent information leakage. When applied at inference time, noise is directly added to model intermediates. In contrast, when applied during training, perturbation is incorporated into intermediates or gradients at each training iteration. While SanText[47], CusText[4] and RanText [35] applies token-level perturbation to hinder inversion. **Learning-based defenses**, such as Mutual Information Defense(MID) [55], Adversarial Training(AT) [24, 40] and TextObfuscator(TO) [51], generally apply a robust training prototype, often with relevant loss regularizers and additional defense models, aiming to divert model representations or gradients leak less information about the privacy target. When applied at inference time, the defense models require prior defense training before being integrated into the SL-LLM system. While applied during training, the defense model is jointly trained with the SL-LLM system. Among the defenses mentioned, TO, SanText, RanText, CusText, and SnD are only designed to defend MIA. Detailed defense method and relevant hyper-parameters are described in Table 5 and Appendix A.

*Evaluation Metrics.* In VFLAIR-LLM, we use various metrics following[52] to assess LLM ability and relevant attack and defense performance.

*Main Task Performance(MP)* refers to the final prediction performance of the SL-LLM system. For Classification tasks(e.g. SST2), MP is defined as the model prediction accuracy. For Regression tasks(e.g. STS-B), MP is the Pearson correlation score. For Text-span based Question Answering tasks(e.g. SQuAD), we take the exact match score as MP. For simple next token prediction tasks(e.g. Lambada), MP is the token prediction accuracy. For QA generation datasets(e.g. Alpaca), we use the Rouge score as MP. For code generation(e.g. CodeAlpaca), we use CodeBLEU[28]. For math tasks(e.g. GSM8K), MP is the problem solving accuracy. *Attack Performance(AP)* refers to the attack success rate. For MIA, AP refers to the recall rate of the recovered texts compared with ground-truth texts. For LIA, it refers to the label recovery accuracy.

*Defense Capability Score (DCS)* is a comprehensive metric for assessing defense effectiveness against a specific attack, considering both the MP and AP as calculated in Equation (1). By default, we set $\beta$ = 0.5 in this paper. *A higher DCS value signifies a superior privacy-utility balance attained by the defense mechanism. Type-level Defense Capability Score (T-DCS)* is the weighted average of DCS for a specific type of attack $j$ (i.e. LIA or MIA), measuring a defense strategy's effectiveness against that attack type as described in Equation (2), where $I_j$ is all the attacks in attack type $j$. In this work, we mainly use $T$-$DCS_{\text{MIA/LIA}}$ and attach equal weight to all attack methods in an attack type. *Comprehensive Defense Capability Score (C-DCS)* is the weighted average of $T$-$DCS$ on various attack types as described in Equation (3), representing the general performance of a defense strategy. $\mathcal{A}$ is all attack types considered. We attach equal weight to all attack types in this research. *DCS Gap($\Delta DCS$)*, is defined as the DCS difference between different methods. In this paper, we mainly evaluate the Full-Vanilla and Full-LoRA fine-tuning in Section 5.3, using $\Delta DCS = DCS_{\text{LoRA}} - DCS_{\text{Vanilla}}$.

$$DCS = \frac{1}{1 + \sqrt{(1-\beta)(AP - AP^*)^2 + \beta(MP - MP^*)^2}} \quad (1)$$

$$T\text{-}DCS_j = \frac{1}{I_j} \sum_{i=1}^{I_j} DCS_i. \quad (2)$$

$$C\text{-}DCS = \sum_{j \in \mathcal{A}} w_j T\text{-}DCS_j, \text{ with } \sum_{j \in \mathcal{A}} w_j = 1.0. \quad (3)$$

## 4 Experiment Settings

Table 7: Evaluated Attacks and Defenses Settings.

| Attack | | Defense | | Deployment | Evaluation Pipeline |
|---|---|---|---|---|---|
| MIA | LIA | Perturbation based | Learning based | | |
| VMI[8] RMI[32] BiSR[3] |  | DP[24],SP[1] SanText[47] CusText[4] RanText[35] SnD [20] | TO[51],MID[55] AT[24] | At Model Head (Figure 4a) | Inference (HB SL-LLM) |
|  | BLI[53] NS[18] | DP[24],SP[1] | MID[55],AT[24] | At Model Tail (Figure 4b) | Fine-tune (HBT SL-LLM) |
| VMI[8] RMI[32] BiSR[3] | BLI[53] NS[18] | DP[24],SP[1] | MID[55],AT[24] | At Both Model Head and Tail (Figures 4a and 4b) | Fine-tune (HBT SL-LLM) |

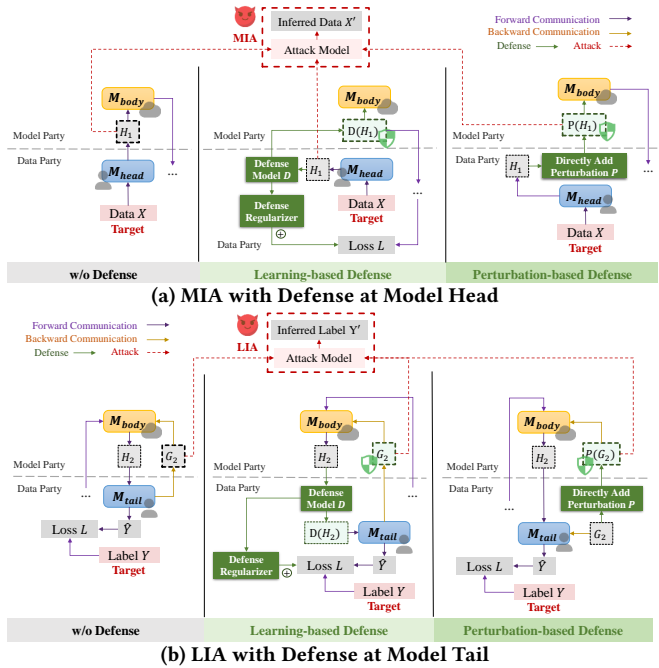

Figure 4: Attacks and Defenses in SL-LLM.
(a) MIA with Defense at Model Head
(b) LIA with Defense at Model Tail

Table 6: SL-LLM Fine-tuning Results

| MP | HB | | | | HBT | | | |
|---|---|---|---|---|---|---|---|---|
| | FL | FV | LL | LV | FL | FV | LL | LV |
| **SST2-Bert** | 0.920±0.001 | 0.905±0.010 | 0.920±0.002 | 0.916±0.006 | 0.919±0.001 | 0.901±0.013 | 0.919±0.003 | 0.915±0.007 |
| **SQuAD-Bert** | 0.731±0.001 | 0.687±0.005 | 0.434±0.002 | 0.705±0.003 | 0.729±0.002 | 0.697±0.010 | 0.708±0.002 | 0.728±0.002 |
| **Lambada-GPT2** | 0.606±0.012 | 0.654±0.002 | 0.566±0.044 | 0.618±0.001 | 0.605±0.007 | 0.653±0.002 | 0.592±0.022 | 0.636±0.001 |

*FL: Full-LoRA  FV: Full-Vanilla  LL: Local-LoRA  LV: Local-Vanilla.

In this section, we provide an overview of the experimental settings in Section 5. Each experiment is tested and averaged on 5 seeds. More detailed configurations are presented in Appendix B.

*Datasets and Models.* We perform the benchmark across various tasks and LLMs, covering 3 classification tasks: SST2-Bert, CoLA-Bert, and Yelp-Bert, and 3 generation tasks: Lamabda-GPT2, Alpaca-GPT2, GSM8K-Mistral, and CodeAlpaca-CodeLlama. We list the detailed configuration for each task in Appendix B.1.

*Attacks and Defenses.* 9 defense methods, 3 MIA and 2 LIA are included in our benchmark as summarized in Tables 4, 5 and 7. For each defense, we comprehensively evaluate different defense hyperparameters described in Table 5, scanning through various defense strengths. We list the detailed attack and defense settings in Appendices B.3 and B.4.

MIA is tested by inverting the training data samples under the final epoch system checkpoint during fine-tuning, while by inverting the test data samples during inference. LIA is tested by inverting batch labels using first epoch gradients following [52] during training. Each attack is evaluated separately, and then MIA and LIA are jointly implemented to evaluate the impact of collaborative defense on both attacks. (termed "MIA-LIA").

*Fine-tuning Strategies.* Four SL-LLM fine-tuning strategies: Full-Vanilla, Full-LoRA, Local-Vanilla, and Local-LoRA as defined in Section 3.2 are evaluated. **Full-Vanilla** refers to fine-tuning all model segments in a vanilla fine-tuning strategy, while **Full-LoRA** fine-tunes all model segments with LoRA. **Local-Vanilla** refers to fine-tuning only the data party's local model segments trainable with vanilla fine-tuning, while **Local-LoRA** uses LoRA to fine-tune the local model segments.

## 5 Experiment Results
## 5.1 SL-LLM Fine-tuning Results.

We evaluate 4 fine-tuning strategies across 3 tasks as presented in Table 6. An early-stop strategy is employed to mark convergence. Detailed experiment settings are listed in Appendix B.2. We notice that, under both Full and Local tuning, using LoRA significantly cuts training time and convergence epochs. For smaller language models (e.g., BERT), LoRA achieves comparable results to Vanilla fine-tuning. However, with larger models (e.g., GPT-2), it results in reduced final accuracy. Full fine-tuning attains better MP than Local fine-tuning across all datasets but requires longer time to reach convergence. Under Local fine-tuning, HBT achieves higher MP than HT, as it fine-tunes a larger set of the model parameters. Both partition configurations yield comparable results under Full-Vanilla strategy.

## 5.2 Attacks and Defenses Benchmark

We apply 9 defense methods deployed on the model head to defend against 3 MIA under HT SL-LLM as presented in Figures 5 and 10. We apply 4 defense methods on the model tail to defend against 2 LIAs under HBT SL-LLM as presented in Figures 7a and 7b. In Figures 6a and 6b, we evaluate 4 defenses against MIA and LIA together by deploying them on both model head and tail.

We present the MP and AP of each attack-defense pair on a 2D **MP-AP graph** to demonstrate overall defense performance. X-axis represents MP while Y-axis indicates AP. Generally, dots closer to the bottom-right achieve higher MP and lower AP, thereby higher DCS, offering a better privacy-utility trade-off.

Analyzing the aforementioned results, we can draw the following conclusions:

**MIA and LIA pose great threats to SL-LLM.** Comparing the black squares illustrating results without defenses in Figures 5 and 7, BiSR, VMI, BLI, and NS achieve high attack accuracy (AP) on most tasks (e.g. > 0.6). VMI and RMI exhibit significantly lower AP on complex tasks like Alpaca and GSM8K in Figures 5b to 5d compared to simpler SST2 task in Figure 5a.

**Privacy-utility trade-off on MP and AP.** In most AP-MP graphs, for each defense, smaller dots, which means defenses with weaker strength, are located to the higher right of larger ones, indicating that as the defense gets stronger, both MP and AP become lower, indicating a trade-off between privacy and utility.

**Learning based methods generally outperform perturbation based methods.** In Figure 5, inference-time perturbation methods suffer significant MP decay with stronger defense strength. Learning-based methods achieve a better MP-AP trade-off, likely because the learning phase can adjust model representation to avoid excessive deviation. Among all defenses, MID showcases superior performance against most attacks, evidenced by its position at the lower-right corner compared to other defenses in Figures 5 to 7 and its leading DCS ranking in Table 8. Although AT performs well in Figure 5, its MP collapses in Figure 6b, indicating potential instability when applied at the model tail. TO matches MID and AT on simple classification (Figure 5a) but underperforms on complex generation tasks(Figures 5b to 5d). Unlike MID and AT, which train a defense model with loss regularizers, TO combines a cluster loss regularizer and random perturbation during training. This dual approach may pose greater challenges when fine-tuning larger LLMs, likely explaining TO's lower performance on complex generation tasks.

**Token-wise perturbation(RanText/CusText/SanText) vs embedding-wise perturbation(DP/SP).** As shown in Figure 5, token-wise perturbation defenses(SanText, RanText, CusText) outperform embedding-wise perturbation(DP, SP) on simple tasks like SST2. However, they fall behind in complex generation tasks like

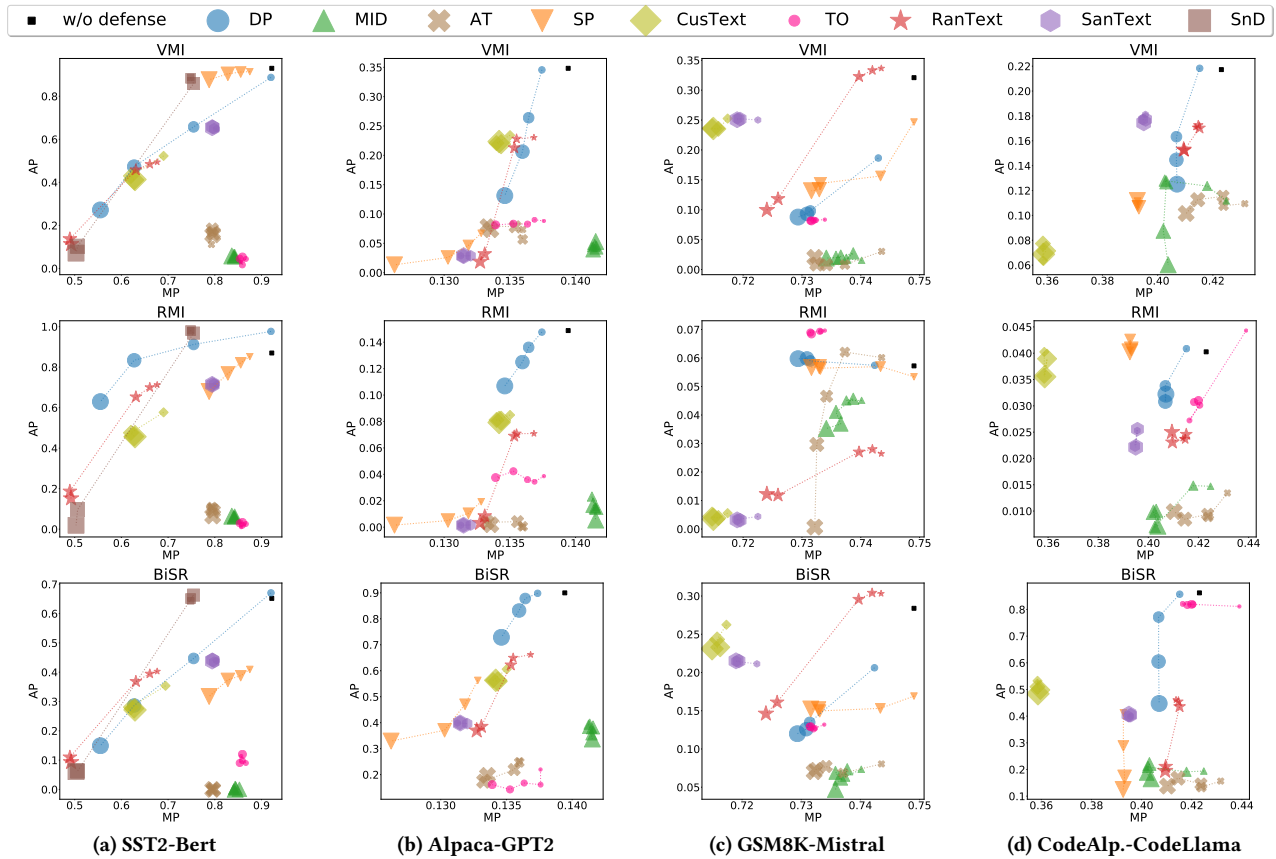

Figure 5: MP-AP results for defending MIA with defense at Model Head under HT SL-LLM. Dot size represents the defense strength, with detailed defense parameters provided in Table 5.

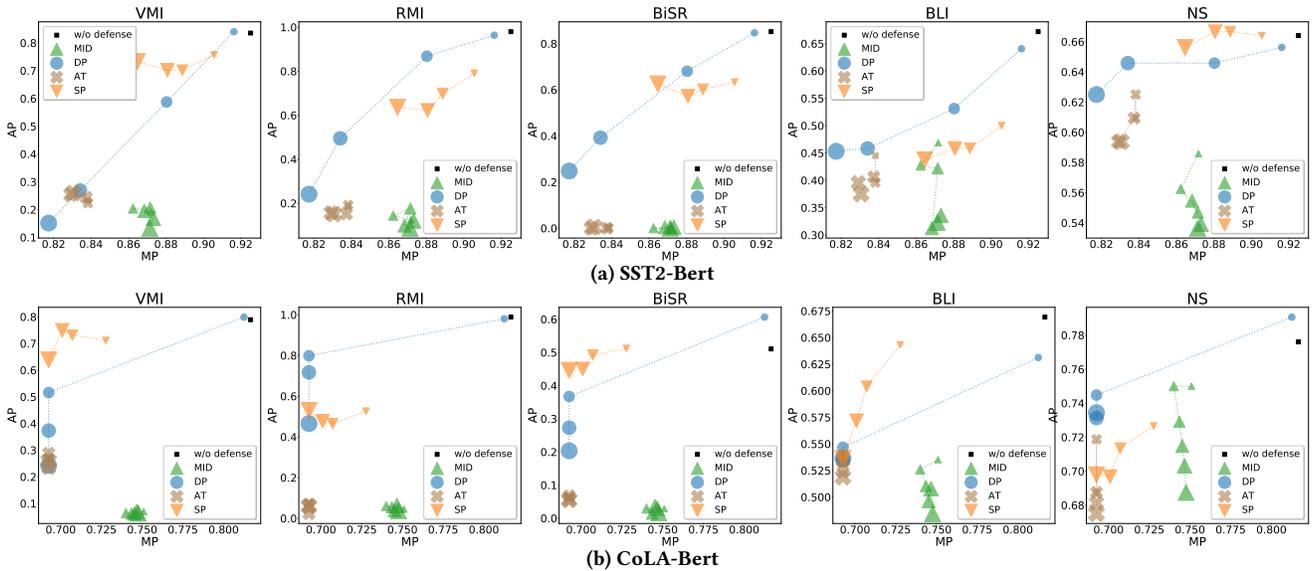

Figure 6: MP-AP results for defending MIA and LIA with defense at both Model Head and Tail under HBT SL-LLM[Full-LoRA fine-tuning strategy]. Dot size represents the defense strength, with detailed defense parameters provided in Table 5.

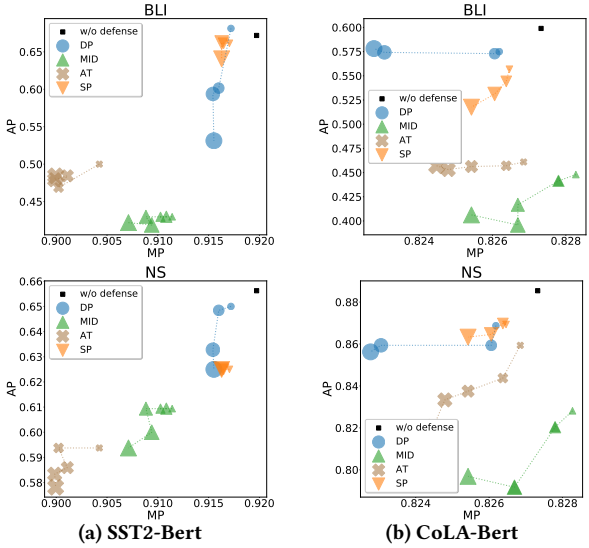

Figure 7: MP-AP results for defending LIA with defense at Model Tail under HBT SL-LLM [Full-LoRA fine-tuning strategy, SST2-Bert].

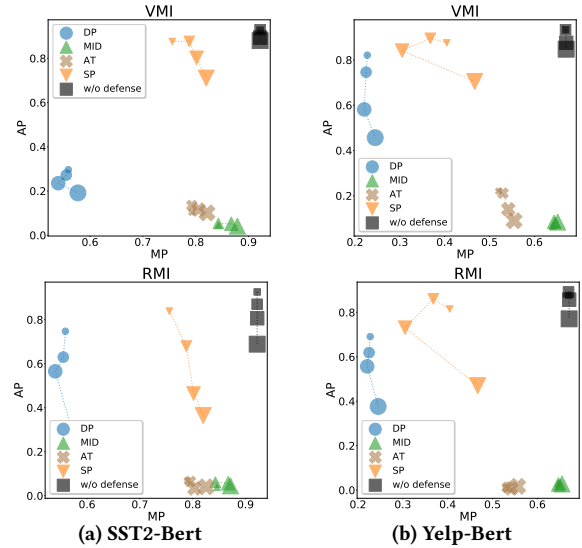

Figure 8: $n_{head}$ Ablation for MIA with defense at Model Head. Dot size represents the scale of $n_{head}$, with the smallest dot representing $n_{head} = 2$ and the largest dot representing $n_{head} = 5$.

Table 8: DCS Ranking. For defending MIA and LIA with defense at both Model Head and Tail under HBT SL-LLM [Full-LoRA fine-tuning strategy].

(a) SST2-Bert

| Defense Name | Defense Parameter | $T\text{-}DCS_{LIA}$ | $T\text{-}DCS_{MIA}$ | $C\text{-}DCS$ | Ranking |
|---|---|---|---|---|---|
| MID | 0.5 | **0.7680** | 0.9347 | 0.8513 | 1 |
| MID | 0.1 | 0.7647 | 0.9230 | 0.8438 | 2 |
| MID | 0.01 | 0.7669 | 0.9204 | 0.8437 | 3 |
| MID | 0.001 | 0.7445 | 0.9055 | 0.8250 | 4 |
| MID | 0.0001 | 0.7402 | 0.9088 | 0.8245 | 5 |
| MID | 1e-05 | 0.7280 | 0.9175 | 0.8228 | 6 |
| AT | 0.001 | 0.7435 | 0.8873 | 0.8154 | 7 |
| AT | 0.1 | 0.7348 | 0.8922 | 0.8135 | 8 |
| AT | 0.01 | 0.7390 | 0.8868 | 0.8129 | 9 |
| AT | 1 | 0.7344 | 0.8892 | 0.8118 | 10 |
| AT | 5 | 0.7269 | 0.8885 | 0.8077 | 11 |
| DP | 50 | 0.7213 | 0.8553 | 0.7883 | 12 |
| DP | 70 | 0.7180 | 0.7829 | 0.7504 | 13 |
| SP | 97 | 0.7168 | 0.6910 | 0.7039 | 14 |
| SP | 98 | 0.7221 | 0.6797 | 0.7009 | 15 |
| SP | 96 | 0.7170 | 0.6794 | 0.6982 | 16 |
| DP | 100 | 0.7061 | 0.6666 | 0.6864 | 17 |
| SP | 95 | 0.7095 | 0.6610 | 0.6853 | 18 |
| DP | 500 | 0.6856 | 0.6158 | 0.6507 | 19 |

(b) CoLA-Bert

| Defense Name | Defense Parameter | $T\text{-}DCS_{LIA}$ | $T\text{-}DCS_{MIA}$ | $C\text{-}DCS$ | Ranking |
|---|---|---|---|---|---|
| MID | 0.5 | **0.7074** | 0.9460 | 0.8267 | 1 |
| MID | 0.01 | 0.7007 | 0.9431 | 0.8219 | 2 |
| MID | 0.1 | 0.7004 | 0.9403 | 0.8204 | 3 |
| MID | 1e-05 | 0.6883 | 0.9450 | 0.8167 | 4 |
| MID | 0.001 | 0.6922 | 0.9409 | 0.8166 | 5 |
| MID | 0.0001 | 0.6895 | 0.9382 | 0.8139 | 6 |
| AT | 0.001 | 0.7023 | 0.8852 | 0.7938 | 7 |
| AT | 0.01 | 0.6959 | 0.8886 | 0.7923 | 8 |
| AT | 0.1 | 0.6969 | 0.8819 | 0.7894 | 9 |
| AT | 1 | 0.6893 | 0.8813 | 0.7853 | 10 |
| AT | 5 | 0.6879 | 0.8818 | 0.7849 | 11 |
| DP | 50 | 0.6874 | 0.8134 | 0.7504 | 12 |
| DP | 70 | 0.6882 | 0.7559 | 0.7221 | 13 |
| SP | 97 | 0.6938 | 0.7155 | 0.7047 | 14 |
| SP | 98 | 0.7006 | 0.7043 | 0.7024 | 15 |
| DP | 100 | 0.6840 | 0.7161 | 0.7001 | 16 |
| SP | 96 | 0.6862 | 0.7137 | 0.6999 | 17 |
| SP | 95 | 0.6807 | 0.7093 | 0.6950 | 18 |
| DP | 500 | 0.6664 | 0.6432 | 0.6548 | 19 |

Alpaca/GSM8K/CodeAlpaca, where detailed input information is crucial for accurate outputs. Token replacement significantly disrupts the input, causing substantial MP loss. In contrast, such disruptions have a less pronounced effect on simple classification tasks. Among token-wise perturbation methods(RanText, CusText, SanText), RanText performs on par with the others on BERT (Figure 5a) but outperforms them on larger language models (Figures 5b to 5d).

**MIA-LIA vs LIA vs MIA.** Comparing results of MIA-LIA(see Figure 6a) and MIA (see Figure 5a), we observe that *perturbation-based defenses applied at training time(Figure 6a) exhibit milder MP decay*. Specifically, a higher MP is reached in Figure 6a than in Figure 5a) especially at lower AP range, suggesting that injecting training-time perturbation for defending MIA and LIA attacks altogether improves MP preservation. On the other hand, learning-based defenses achieve comparable performance in both scenarios. The comparison between MIA-LIA (Figure 6) and LIA (Figure 7) reveals that *MP deteriorates when perturbations are applied at both the model head and tail(Figure 6) versus at the tail alone(Figure 7)*. Perturbing both intermediates and gradients during training leads to increased MP loss.

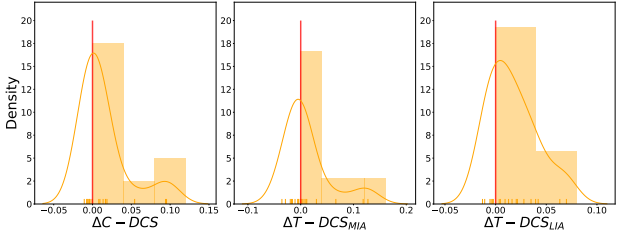

Figure 9: DCS Gap Distribution. For Defending MIA and LIA with defense at both Model Head and Tail under HBT SL-LLM [SST2-Bert]. Here, $\Delta DCS = DCS_{\text{Full-LoRA}} - DCS_{\text{Full-Vanilla}}$, positive $\Delta DCS$ indicate Full-LoRA outperform Full-Vanilla.

Table 9: Average DCS Gap. For Defending MIA and LIA with defense at both Model Head and Tail under HBT SL-LLM [SST2-Bert].

|  | $\Delta C$-DCS | $\Delta T$-$DCS_{\text{MIA}}$ | $\Delta T$-$DCS_{\text{LIA}}$ |
| --- | --- | --- | --- |
| **Overall Average** | 0.0141 | 0.0140 | 0.0142 |
| MID Average | 0.0071 | 0.0016 | 0.0127 |
| AT Average | 0.0463 | 0.0501 | 0.0425 |
| SP Average | 0.0060 | 0.0109 | 0.0011 |
| DP Average | -0.0076 | -0.0094 | -0.0058 |

## 5.3 Further Ablation Studies

**Larger model head achieves better privacy-utility trade-off but demands more local resources.** To understand the impact of model head size, we evaluate 4 HT SL-LLM settings with the number of model head decoders $n_{head}$ ranging from 2 to 5. 4 defenses are tested against MIA on 2 classification tasks(SST2/Yelp) as shown in Figure 8. For each defense, we choose hyperparameters that yield the best DCS. As shown in Figure 8, larger dots tend to position towards the lower right, suggesting an enhanced privacy-utility trade-off introduced by larger $n_{head}$. The trend is more pronounced for perturbation-based defenses, while learning-based ones are less affected. However, hosting a larger model head demands more local resources, introducing another trade-off in the design and deployment of SL-LLM systems.

**SL-LLM with LoRA fine-tuning is more robust against privacy attacks.** To explore the impact of fine-tuning strategies, we evaluate **Full-Vanilla** and **Full-LoRA** fine-tuning on SST2 under HBT SL-LLM, applying defense at both model head and tail. Let $\Delta DCS$ denote the DCS Gap between **Full-LoRA** and **Full-Vanilla** methods, where a positive value indicates better privacy-utility performance achieved through the use of LoRA. Similarly, $\Delta C$-$DCS$ is defined as the C-DCS Gap between **Full-LoRA** and **Full-Vanilla** methods, while $\Delta T$-$DCS_{\text{MIA}}$, $\Delta T$-$DCS_{\text{LIA}}$ follow similar definitions.

In Figure 9, we present the histogram of $\Delta DCS$ for the tested defenses. The average DCS gap of each defense type is summarized in Table 9. As most orange histograms appear at the right of the vertical line marking 0.0 in Figure 9 and overall average DCS gap above 0 in Table 9, we conclude that SL-LLM with LoRA is more resistant to privacy attacks than vanilla training.

## 6 Conclusions

In this work, we introduce **VFLAIR-LLM**, a lightweight and extensible SL-LLM framework that incorporates fundamental inference and fine-tuning pipelines within two LLM partition settings. The framework incorporates a broad spectrum of LLM types, working modes, attack and defense algorithms, supporting diverse tasks and datasets. Additionally, we provide a thorough benchmark on SL-LLM privacy algorithms, delivering practical insights on various attack and defense strategies, which serve as valuable guidance for users to select appropriate strategies in real-world applications. While VFLAIR-LLM provides a versatile framework, further research is still needed for the acceleration of SL-LLM inference and fine-tuning.

## 7 Acknowledgment

This work was supported by the National Key R&D Program of China under Grant No.2022ZD0160504, and Wuxi Innovation Center of Tsinghua AIR, under Grant A20240103.

**Algorithm 2** A Normal HBT SL-LLM Training Procedure.

**Input**: learning rates $\eta$, LoRA strategy $LoRA()$
**Output**: $\mathcal{M}_{head}, \mathcal{M}_{body}, \mathcal{M}_{tail}$.

1: Model Party initialize $\mathcal{M}_{body}$; Data Party initialize $\mathcal{M}_{head}$, $\mathcal{M}_{tail}$.
2: **for** $i \in [head, body, tail]$ **do**
3:     **if** $\mathcal{M}_i$ is trainable **then**
4:         **if** Use LoRA Strategy **then** $\mathcal{M}_i \leftarrow LoRA(\mathcal{M}_i)$
5:         **else**
            Freeze $\mathcal{M}_i$
6:     **end if**
7: **end for**
8: **for** each iteration $j = 1, 2, ...$ **do**
9:     Randomly sample a mini-batch of samples $\{x, y\} \subset \mathcal{D}$ of size n;
10:    Data Party computes $H_{1,k} = \mathcal{M}_{head}(x_k)$ and sends it to model party;
11:    Model Party computes $H_{2,k} = \mathcal{M}_{body}(H_{1,k})$ and sends it to data party;
12:    Data Party computes the prediction $\hat{y}_k = \mathcal{M}_{tail}(H_{2,k})$;
13:    Data party computes the loss $\mathcal{L} = \frac{1}{n}\ell(y, \hat{y})$ and the gradient $G_1 = \frac{\partial \mathcal{L}}{\partial H_2}$, then sends the gradient to model party;
14:    Data party updates $\mathcal{M}_{tail}^{j+1} = \mathcal{M}_{tail}^j - \eta_1 \frac{\partial \mathcal{L}}{\partial \mathcal{M}_{tail}}$;
15:    Model party updates $\mathcal{M}_{body}^{j+1} = \mathcal{M}_{body}^j - \eta_1 G_1 \frac{\partial H_2}{\partial \mathcal{M}_{body}}$;
16:    Model Party computes the gradient $G_2 = G_1 \frac{\partial H_2}{\partial H_1}$ and sends it to data party;
17:    Data party updates $\mathcal{M}_{head}^{j+1} = \mathcal{M}_{head}^j - \eta_1 G_2 \frac{\partial H_1}{\partial \mathcal{M}_{head}}$;
18: **end for**=0

**Algorithm 1** A Normal HT SL-LLM Training Procedure.

**Input**: Learning rates $\eta$, LoRA strategy $LoRA()$.
**Output**: $\mathcal{M}_{head}, \mathcal{M}_{tail}$.

1: Model Party initialize $\mathcal{M}_{tail}$; Data Party initialize $\mathcal{M}_{head}$.
2: **for** $i \in [head, tail]$ **do**
3:     **if** $\mathcal{M}_i$ is trainable **then**
4:         **if** Use LoRA Strategy **then** $\mathcal{M}_i \leftarrow LoRA(\mathcal{M}_i)$
5:         **else**
            Freeze $\mathcal{M}_i$
6:     **end if**
7: **end for**
8: **for** each iteration $j = 1, 2, ...$ **do**
9:     Randomly sample a mini-batch of samples $\{x, y\} \subset \mathcal{D}$ of size n;
10:    Data Party computes $H_k = \mathcal{M}_{head}(x_k)$ and sends it to model party;
11:    Model Party computes the prediction $\hat{y}_k = \mathcal{M}_{tail}(H_k)$ and sends it to data party;
12:    Data party computes the loss $\mathcal{L} = \frac{1}{n}\ell(y, \hat{y})$ and the gradient $G_1 = \frac{\partial \mathcal{L}}{\partial \hat{y}}$, then sends the gradient to model party;
13:    Model party updates $\mathcal{M}_{tail}^{j+1} = \mathcal{M}_{tail}^j - \eta_1 G_1 \frac{\partial \hat{y}}{\partial \mathcal{M}_{tail}}$;
14:    Model Party computes the gradient $G_2 = G_1 \frac{\partial \hat{y}}{\partial H}$ and sends it to data party;
15:    Data party updates $\mathcal{M}_{head}^{j+1} = \mathcal{M}_{head}^j - \eta_1 G_2 \frac{\partial H}{\partial \mathcal{M}_{head}}$;
16: **end for**=0

## A Supported Defenses
### A.1 Perturbation-based Defenses

**Differential Privacy(DP) [7, 24]** is implemented by clipping and adding noise to intermediate results or gradients. Larger $\epsilon$ in the added Laplace noise $Lap(\Delta f/\epsilon)$ indicates stronger perturbation and defense, where $\Delta f$ denotes the l1-sensitivity [24] of the LLM. **Sparsification(SP) [1, 9, 54]** is implemented by dropping elements in tensors that are close to 0. Sparsification rate $r$ is the percent of sparsified coordinates in tensors. Larger $r$ indicates stronger perturbation and therefore stronger defense. **SanText [47]/CusText [4]/RanText [35]** all employ token-wise perturbation based on an MLDP mechanism, where $\epsilon$ controls the DP noise level. A larger $\epsilon$ corresponds to greater MLDP noise, indicating stronger defense. Specifically, SanText replaces a portion of tokens with one close in terms of embedding distance from a word adjacency list. While CusText perturbs all words in a sentence and uses a smaller word adjacency list. RanText introduces a random adjacency list mechanism and samples perturbed words via MLDP to perturb documents. In **Split-N-Denoise(SnD) [20]**, the data party first perturbs the intermediate embedding via a DP-based privatization module. The received noised embedding from model party is subsequently denoised using a pre-trained denoising model, offering inference-time defense for classification tasks. Larger $\eta$ corresponds to weaker noise in the perturbation module and therefore weaker defense.

### A.2 Learning-based Defenses

**Mutual Information Defense(MID) [40, 55]** introduces a bottleneck known as the Mutual Information (MI) model into the data party's model. This MI model is trained with a mutual information loss regularizer, which is the mutual information between the privacy target and the intermediate tensor obtained by potential attackers, steering the intermediates away from revealing the privacy target. This defense is primarily designed for computer vision (CV) tasks but can be extended to LLM applications as its mechanism is independent of the model architecture. Regularizer Strength $\lambda$ controls the weight of the MI regularizer in the training loss. Larger $\lambda$ attaches more importance to minimizing the

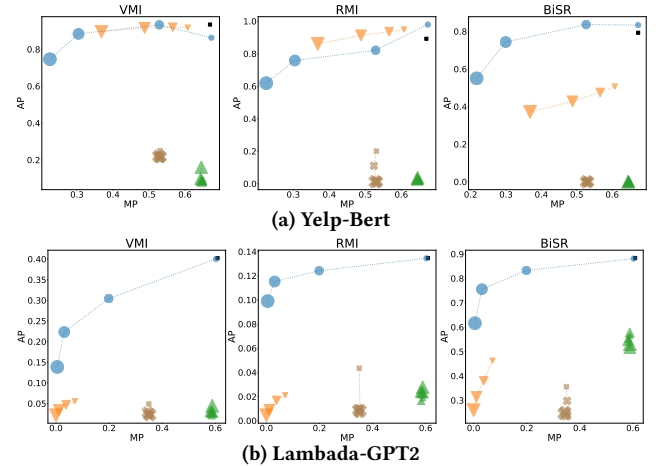

Figure 10: MP-AP results for defending MIA with defense at Model Head under HT SL-LLM. Dot size represents the defense strength.

Table 10: Efficiency Evaluation for Distributed/Standalone Deployment

| | Throughput (token/s) | Communication Avg.(kb/token) | Communication Total(MB) |
|---|---|---|---|
| Llama3-8B(Std.) | 23.27 | / | / |
| Llama3-70B(Std.) | 9.66 | / | / |
| Llama3-8B(Dist.) | 15.41 | 16 | 340 |
| Llama3-70B(Dist.) | 6.97 | 32 | 1473 |

Std.:standalone, Dist.:distributed.

MI regularizer, indicating stronger defense. **Adversarial Training(AT) [24]** is a widely used learning-based defense strategy. In AT, a simulated adversary model $A_\phi$ is trained jointly with a Privacy Preserving Mapping[29] $D$ trying to minimize the effectiveness of the imagined adversary. A mapping distance $\lambda||D_\theta(H) - H||^2$ is added as a utility regularizer, constraining the embedding distortion. The overall training can be presented as a minmax problem: $\min(L_f(Y, \hat{Y}) + L_f(A_\phi(D_\theta(H)), X) + \lambda||D_\theta(H) - H||_2)$. Larger regularizer Strength $\lambda$ indicates more emphasis on utility and weaker defense. **TextObfuscator(TO) [51]** fine-tune the whole LLM system with word representation obfuscation and a cluster loss regularizer. It can only defend MIA and has proved efficient on RoBERTa for classification tasks. A larger cluster number $n_{\text{cluster}}$, indicates more precise word clustering, which results in reduced perturbation and therefore weaker defense strength.

## B Detailed Experimental Settings

### B.1 Task and Model Configurations

**SST-2**, **CoLA**, **Yelp** and **SQuAD** are tested on open-sourced bert-based models available at https://huggingface.co/textattack/bert-base-uncased-SST-2, https://huggingface.co/Shunian/yelp_review_classification and https://huggingface.co/google-bert/bert-large-uncased-whole-word-masking-finetuned-squad. In HT setting, $n_{head} = 3$. In HBT setting, $n_{head} = 3$ and $n_{tail} = 3$. **Lambada** is tested on GPT2 model[26]. In HT setting, $n_{head} = 2$. In HBT setting, $n_{head} = 2$ and $n_{tail} = 2$. **Alpaca** is tested on an open-sourced GPT2 model available at https://huggingface.co/vicgalle/gpt2-alpaca. In HT setting, $n_{head} = 2$. **GSM8K** is tested on an open-sourced Mistral-7B model[46] available at https://huggingface.co/meta-math/MetaMath-Mistral-7B. In HT setting, $n_{head} = 2$. **CodeAlpaca** is tested on the open-sourced CodeLlama-7B model available at https://huggingface.co/codellama/CodeLlama-7b-hf. In HT setting, $n_{head} = 2$.

### B.2 Fine-tuning Settings

In Section 5.1, we set training bs(batch-size) to 128/32/16 and lr (learning rate) to 1e-4/5e-5/1e-5 for SST2, SQuAD and Lambada respectively. Convergence is marked with an early-stop strategy. For LoRA fine-tuning, we set $r = 4$, $\alpha = 32$, and dropout rate=0.1, which remains the same in all other LoRA experiments.

### B.3 Attack Settings

**Model Inversion Attack Settings:** For SST2, CoLA, Yelp, Lambada, Alpaca, GSM8K, CodeAlpaca, we randomly sample 100/100/1000/200/100/100/100 samples from the targeting dataset for evaluating inversion attacks. In VMI, attack training epoch is set to 400/100 with a lr of 0.001/0.01 for Lambada/other datasets. In RMI, attack training epoch is set to 300/500 with lr of 0.005 for Lambada/other datasets. We set temperature to 0.5 following [32]. In BiSR, we randomly select 300/100 training samples for the first noise-aware pre-training phase for Alpaca/other datasets, where expert epoch=20, gate epoch=15 and full epoch=4 following [3]. In the subsequent data reconstruction phase, attack epoch is set to 100 with lr=0.01.
**Label Inference Attack Settings:** In BLI, we set attack training epoch to 500 with a lr of 0.05 for all tested datasets, while NS does not require additional hyper-parameter setting.

### B.4 Defense Settings

**Apply Defense at Model Head in HT SL-LLM:** In Figure 5, perturbation are injected directly into SL-LLM inference. In SanText, percentage of perturbed sensitive words is set to 0.5. In CusText, top-k is set to 20% for SST2, Alpaca and CodeAlpaca, 100% for GSM8K. In SnD, denoise model is pretrained on the training set for 16 epochs, with lr=0.0001 and bs=12. $\eta_{train}$ is set to 100 following [20]. MID and AT requires SL-LLM training to train its defense model. For SST2/Yelp/Alpaca, we set bs=128/64/12 and lr=1e-4 with padding length of 70/384/256. For Lambada, we generate the training samples of length 512. bs=32 and lr=1e-3. For GSM8K, we set bs=16, lr=0.001, with padding length of 128. In TO, for SST2, we apply a Full-Vanilla strategy and set bs=128,lr=0.0001, $w_{away} = 0.5, w_{close} = 0.1, \epsilon = 1$. For Alpaca, we apply a Full-Vanilla strategy and set bs=16,lr=0.0001, $w_{away} = 0.5, w_{close} = 0.1, \epsilon = 2$. For GSM8K, we apply a Full-LoRA strategy and set bs=8,lr=1e-8, $\epsilon = 2$. We set $w_{away} = 0.5, w_{close} = 0.1$ for all datasets.
**Apply Defense at Model Tail in HBT SL-LLM:** In Figures 7a and 7b, we apply a Full-LoRA fine-tuning strategy. For SST2, we use bs=16, and LLM lr=0.0001. In MID and AT, the defense model lr is 0.0001. For CoLA, bs=32, lr=0.0001. Defense model lr is set to 0.0001 for MID and AT.
**Apply Defense at Both Model Head and Tail in HBT SL-LLM:** In Figures 6a and 6b, we apply a Full-LoRA fine-tuning strategy. For SST2/CoLA, main lr=1e-4 and bs=16/32. MID and AT's defense model lr is set to 0.0001. When applying Full-Vanilla in Table 9 for SST-2, main LLM lr is altered to 1e-5.

## C Distributed Deployment Result of SL-LLM

We evaluate the efficiency of distributed vs standalone SL-LLM inference using Llama3-8B and Llama3-70B in Table 10, tested by conducting inference on 100 randomly selected samples from the Alpaca dataset. For distributed deployment, we use 2 Nvidia A100 GPUs on the model party side and 1 Nvidia A10 GPU on the data party side with a bandwidth of 300Mb for communication. For standalone simulation, we use 2 Nvidia A100 GPUs. Both models show significantly lower throughput in distributed mode compared to standalone deployment, suggesting communication efficiency is a bottleneck for distributed deployment which requires further improvement.

## D Additional Experiment Results

Due to space limit, we place additional MIA result on other datasets in Figure 10. We also provide a detailed **user guidance** for VFLAIR-LLM in our code base(see https://github.com/FLAIR-THU/VFLAIR-LLM).